\documentclass{aa}  
\usepackage{graphicx}
\usepackage{rotating}
\usepackage[varg]{txfonts}
\usepackage{hyperref}
\usepackage{natbib}
\bibpunct{(}{)}{;}{a}{}{,}

\begin{document}
\title{The Chemical Evolution of Fluorine in the Bulge\thanks{Based on observations collected at the European Southern Observatory, Chile (ESO programs 71.B-0617(A), 073.B0074(A), and 079.B-0338(A)) and observations obtained at the Gemini Observatory, which is operated by the Association of Universities for Research in Astronomy, Inc., under a cooperative agreement with the NSF on behalf of the Gemini partnership: the National Science Foundation (United States), the National Research Council (Canada), CONICYT (Chile), the Australian Research Council (Australia), CNPq (Brazil), and CONICRT (Argentina), as program GS-2004A-Q-20.}}
\subtitle{High-resolution K-band spectra of giants in three fields}
\author{H.~J\"onsson\inst{1} \and N.~Ryde\inst{1} \and G.~M.~Harper\inst{2} \and K.~Cunha\inst{3} \and M.~Schultheis\inst{4} \and K.~Eriksson\inst{5} \and C.~Kobayashi\inst{6} \and V.~V.~Smith\inst{7} \and M.~Zoccali\inst{8} }
\institute{Lund Observatory, Department of Astronomy and Theoretical Physics, Lund University, Box 43, SE-221 00 Lund, Sweden\\ \email{henrikj@astro.lu.se}\and
School of Physics, Trinity College, Dublin 2, Ireland\and
Observat\'{o}rio Nacional, Rua General Jos\'{e} Cristino, 77, 20921-400 S\~ao Crist\'{o}v\~ao, Rio de Janeiro, RJ, Brazil\and
Observatoire de la Cote d'Azur, Boulevard de l'Observatoire, B.P. 4229, F 06304 NICE Cedex 4, France\and 
Department of Physics and Astronomy, Uppsala University, Box 516, SE-751 20 Uppsala, Sweden \and
Centre for Astrophysics Research, University of Hertfordshire, Hatfield AL10 9AB, United Kingdom\and
National Optical Astronomy Observatory, 950 North Cherry Avenue, Tucson, AZ 85719, USA \and
Instituto de Astrofisica, Pontifica Universidad Catolica de Chile, Casilla 306, Santiago 22, Chile
}
\date{Submitted 2014; accepted 2014}

\abstract
{Possible main formation sites of fluorine in the Universe include AGB stars, the $\nu$-process in Type II supernova, and/or Wolf-Rayet stars. The importance of the Wolf-Rayet stars has theoretically been questioned and they are probably not needed in the modelling of the chemical evolution of fluorine in the solar neighborhood. It has, however, been suggested that Wolf-Rayet stars are indeed needed to explain the chemical evolution of fluorine in the Bulge. The molecular spectral data, needed to determine the fluorine abundance, of the often used HF-molecule has not been presented in a complete and consistent way and has recently been debated in the literature.}
{We intend to determine the trend of the fluorine-oxygen abundance ratio as a function of a metallicity indicator in the Bulge to investigate the possible contribution from Wolf-Rayet stars. Additionally, we present here a consistent HF line list for the K- and L-bands including the often used 23358.33~\AA\,line.}
{High-resolution near-infrared spectra of eight K giants were recorded using the spectrograph CRIRES mounted at VLT. A standard setting was used covering the HF molecular line at 23358.33~\AA. The fluorine abundances were determined using spectral fitting. We have also re-analyzed five previously published Bulge giants observed with the Phoenix spectrograph on Gemini using our new HF molecular data.}
{We find that the fluorine-oxygen abundance in the Bulge probably cannot be explained with chemical evolution models including only AGB-stars and he $\nu$-process in supernovae Type II, i.e. a significant amount of fluorine production in Wolf-Rayet stars is likely needed to explain the fluorine abundance in the Bulge. Concerning the HF line list, we find that a possible reason for the inconsistencies in the literature, with two different excitation energies being used, is  two different definitions of the zero-point energy for the HF molecule and therefore also two accompanying different dissociation energies. Both line lists are correct, as long as the corresponding consistent partition function is used in the spectral synthesis. However, we suspect this has not been the case in several earlier works leading to fluorine-abundances $\sim$0.3 dex too high. We present a line list for the K- and L-bands and an accompanying partition function.}
{}

\keywords{Galaxy: bulge --  Galaxy: evolution -- Stars: abundances -- Infrared: stars}
\maketitle
\section{Introduction} \label{sec:introduction}
From a nucleosynthetic perspective fluorine is a very interesting element and its cosmic origin is truly intriguing.  Its creation and destruction in stellar interiors is very sensitive to the physical conditions \citep[see for example][]{2011ApJ...729...40L}, meaning that observations of fluorine abundances can provide strong constraints to stellar models. It will also be possible to observationally constrain the main stellar nuclear production sites of fluorine in the Universe at different epochs and in different stellar populations. To do this, observations of the chemical evolution of fluorine as a function of metallicity for different stellar populations have to be confronted with model predictions.

Theoretical considerations have offered three main production mechanisms which all should work under prevailing conditions during different phases of stellar evolution. Their relative importance at different stages of evolution and in different stellar populations is only starting to be investigated. The different production sites of \element[][19]{F}, the only stable isotope of fluorine, that have been proposed are:

\begin{itemize}
\item \textit{$\nu$ nucleosynthesis in supernovae Type II (SNe II)}\\
The core collapse of a massive star, following a SN II explosion, leads to a prodigious neutrino flux. In spite of the small cross sections, the large amount of neutrinos gives rise to a significant spallation of \element[][20]{Ne} to \element[][19]{F} \citep{1988Natur.334...45W} in the overlying (neon-rich) shells of the core. \citet{1991NuPhA.527..663H} estimates the total (mu- and tau-) neutrino energy to $E_\nu=3\times 10^{53}$\,erg. \citet{2011ApJ...739L..57K} investigate the importance of this total neutrino energy for the $\nu$ process reactions for the evolution of fluorine in the solar neighborhood. They conclude that the $\nu$ nucleosynthesis should be a major fluorine production mechanism and that its relative contribution is largest for low metallicities.

\item \textit{Thermal-pulsing Asymptotic Giant Branch (TP-AGB) stars}\\ 
Low-mass ($2 \la M/M_{\odot} \la 4$) TP-AGB stars have been suggested to produce fluorine in different burning phases during the thermal pulse stage, by nuclear reaction chains starting from $\element[][14]{N}$ \citep{1992A&A...261..157F,1992A&A...261..164J,2011ApJ...737L...8A,2011ApJ...739L..57K,2010MmSAI..81..998G}. Fluorine is then transported up to the surface by the ${3}^\mathrm{rd}$ dredge-up. Fluorine production in AGB-stars is expected to be accompanied by the slow-neutron capture nucleosynthesis (the s-process), producing elements like Sr, Y, Zr, Nb, Ba, and La \citep[e.g.][]{1998A&A...334..153M,2000A&A...362..599G,2009ApJ...694..971A}. It has been demonstrated observationally that AGB stars do produce fluorine, see for example \citet{1992A&A...261..164J} and \citet{2011ApJ...737L...8A}.

\item \textit{Wolf-Rayet (W-R) stars}\\
\citet{1993nuco.conf..503M,1996LIACo..33...89M,2000A&A...355..176M} suggested that W-R stars could contribute significantly to the galactic fluorine budget. $\element[][19]{F}$ is produced in the convective cores of W-R stars, during the core He-burning phase. Due to a large mass loss caused by a metallicity-dependent, radiatively-driven wind, the destruction of $\element[][19]{F}$ by the ($\alpha$, $p$) reaction is prevented since the convective core shrinks. The fluorine left behind is eventually exposed at the surface as the heavy mass loss strips the star of the outer layers. This mechanism depends on key parameters, such as initial mass, metallicity, and rotational velocity. Fluorine is produced from $\element[][14]{N}$, which means that the more $\element[][14]{N}$ is available the more fluorine is expected. A second metallicity-dependent effect is the metallicity-dependent winds. Both circumstances favor the fluorine production at higher metallicities. \citet{2005A&A...443..243P} show that when incorporating newer yields and including models of rotating W-R stars, the yields from this mechanism are significantly reduced, implying that W-R stars might not be a major contributor of fluorine. However, they conclude that due to large uncertainties in key nuclear-reaction rates and mass-loss rates, the question of the contribution to galactic $\element[][19]{F}$ from W-R stars is still open.
\end{itemize}

Using a semi-analytic multizone chemical-evolution model, \citet{2004MNRAS.354..575R} show for the first time the impact of the AGB and W-R star contributions to the Galactic chemical evolution of fluorine. They show that $\nu$ nucleosynthesis was dominant in the early universe and that AGB stars' significance successively grows. Based on the old yields and non-rotating models, they further show that the contribution of W-R stars is significant for solar and super-solar metallicities, increasing the [F/O] ratio by a factor of two at solar metallicities. Their conclusion is that all three production sites are needed in order to explain the Galactic chemical evolution of fluorine for a range of metallicities.

\citet{2011ApJ...739L..57K} modeled the evolution of fluorine in the solar neighborhood including AGB stars and $\nu$ nucleosynthesis with two different neutrino energies ($E_\nu=3\times 10^{53}$\,erg and $E_\nu=9\times 10^{53}$\,erg). 
Note that the contributions from W-R stars are underestimated in these models, because the elements such as C, N, and possibly F that are newly produced and have been lost via stellar winds before supernova explosions are not included. The models show a good agreement with field stars of higher metallicities. At lower metallicities the models cannot reproduce the observations of \citet{2013ApJ...765...51L}, but still the model that fits best include the $\nu$ process with $E_\nu=3\times 10^{53}$\,erg. 

The abundance of fluorine in stars is difficult to measure due to a paucity of suitable spectral lines. Highly ionized \ion{F}{v} and \ion{F}{vi} lines in the UV have been used by \citet{2005A&A...433..641W} in extremely hot post-AGB stars and a handful of \ion{F}{i} lines between 6800-7800~\AA\,have been used in extreme helium stars and R Coronae Borealis stars \citep{2006ApJ...648L.143P,2008ApJ...674.1068P}. All other studies we are aware of have been made using HF molecular lines in the K-band and mostly the HF($1-0$) R9 line at 23358.329~\AA. 

Relevant for the observations we present in this paper, is the study by \citet{2008ApJ...679L..17C} who present the first 
study of the chemical evolution of fluorine in the Galactic Bulge, by investigating six red giants in Baade's Window (five of these spectra are re-analyzed in this paper). They find that the fluorine to oxygen abundance ratio in the Bulge follows and extends the solar neighborhood trend. The trend at higher metallicities needs other sources of fluorine in addition to the $\nu$ process contribution, which is sufficient at lower metallicities. These are the AGB star and W-R star contributions. By investigating the correlation with abundances of s-process elements, the authors conclude that, for the Bulge, the W-R wind contribution to the fluorine budget should be important and larger than for the Disk. They therefore suggest that W-R stars might have played a vital role in the chemical evolution of the Galactic Bulge.

In this paper, we observationally investigate the chemical evolution of fluorine in the Bulge, by analyzing red giants from three fields. We discuss the relative contributions of the different main nucleosynthetic sites suggested, by comparing with the latest and most updated models for the evolution of fluorine in the Bulge. Our main conclusion is that a significant fluorine production in W-R stars is likely needed to explain the fluorine abundance in the Bulge, meaning that the production in AGB-stars and SNe II is probably not enough.

\section{Observations}
\begin{figure*}[htp]
\centering
\includegraphics[width=180mm]{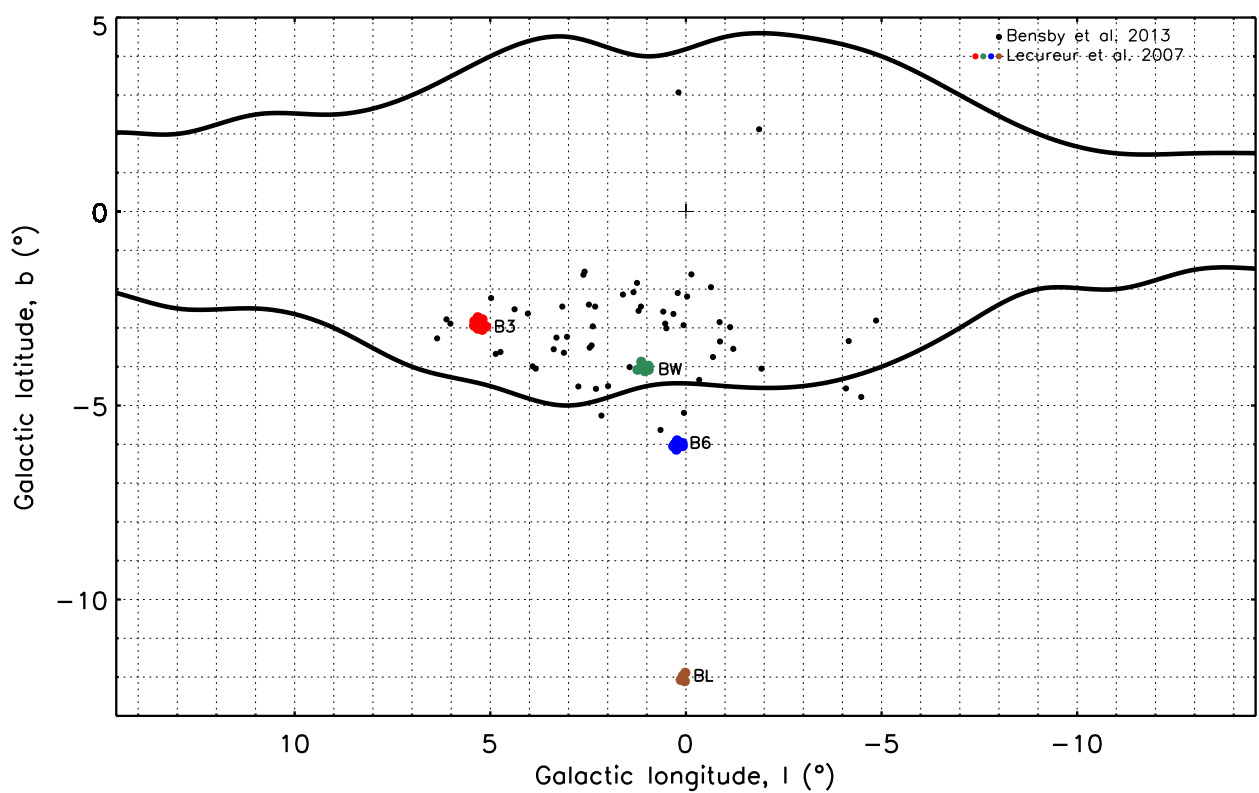}
\caption{Location of the four fields (B3, BW, B6, and BL) of \citet{2007A&A...465..799L} in comparison to the COBE/DIRBE outline of the Galactic Bulge \citep{1994ApJ...425L..81W} and the study of \citet{2013A&A...549A.147B}. Our stellar sample is a subset of the B3-, BW-, and B6-stars. The five re-analyzed stars from \citet{2008ApJ...679L..17C} are in the BW-field.}
\label{fig:bulge_fields}
\end{figure*}

We have observed eight K giants in the galactic Bulge using the spectrometer CRIRES \citep{2004SPIE.5492.1218K,2005hris.conf...15M,2006Msngr.126...32K}, mounted on VLT. 
The K-band observations explored in this paper are, with one exception,  of the same stars as the H-band observations analyzed in \citet{2010A&A...509A..20R}, in turn a sub-sample of the full visual sample used in \citet{2006A&A...457L...1Z}, \citet{2007A&A...465..799L}, and \citet{2013A&A...559A...5B}. The basic data of our stars are listed in Table \ref{tab:basicdata} and the Figure \ref{fig:bulge_fields} shows the location of our three fields (B3, BW and B6) in comparison to the COBE/DIRBE outline of the Galactic Bulge \citep{1994ApJ...425L..81W} and the micro lensed Bulge dwarfs of \citet{2013A&A...549A.147B}.

\begin{table*}[htp]
\caption{Basic data for the observed red giants.}
\begin{tabular}{l r c c c c c c c c c}
\hline
\hline
\multicolumn{1}{c}{Star$^a$} & \multicolumn{1}{c}{OGLE no} & RA (J2000) & Dec (J2000) & $I$ & $V-I$ & $H$ & $K$\\
         &         & (h:m:s)    & (d:am:as)   &     &       & \\
\hline
\textrm{B3-b1} & 132160C4 & 18:08:15.840 & -25:42:09.83 & 16.345 & 2.308 & 11.525 & 11.310\\
\textrm{B3-b7} & 282804C7 & 18:09:16.540 & -25:49:26.08 & 16.355 & 2.304 & 11.614 & 11.351\\
\textrm{B3-b8} & 240083C6 & 18:08:24.602 & -25:48:44.39 & 16.488 & 2.427 & 11.395 & 11.130\\
\textrm{B3-f3} &  95424C3 & 18:08:49.628 & -25:40:36.93 & 16.316 & 2.259 & 11.676 & 11.464\\
\textrm{BW-f6} &   392918 & 18:03:36.890 & -30:07:04.30 & 16.370 & 2.017 & 12.043 & 11.832\\
\textrm{B6-b8} & 108051c7 & 18:09:55.950 & -31:45:46.33 & 16.290 & 2.107 & 11.883 & 11.653\\
\textrm{B6-f1} &  23017c3 & 18:10:04.460 & -31:41:45.31 & 15.960 & 1.941 & 11.914 & 11.671\\
\textrm{B6-f7} & 100047c6 & 18:10:52.300 & -31:46:42.18 & 15.950 & 1.891 & 11.904 & 11.734\\
\hline
\end{tabular}
\label{tab:basicdata}
\tablefoot{\\
\tablefootmark{a}{Using the same naming convention as \citet{2007A&A...465..799L}.}
}
\end{table*}

The stars were observed with the CRIRES-setting 24/-1/i giving a spectral coverage from approximately 23070~\AA\,to 23510~\AA\,and therefore including the HF-line at 23358.33~\AA. The spectral resolution is around R=40000, as determined from narrow telluric lines. The observations were reduced using the CRIRES pipeline and the continua are normalized with the the IRAF task \texttt{continuum}. \label{sec:cont}
Subsequently, the telluric lines plaguing this part of the IR spectra, were carefully removed by dividing the normalized spectra with that of a telluric standard of high signal-to-noise ratio, which we observed in the same setting and reduced in the same way, using the IRAF task \texttt{telluric}. 

The stellar parameters were re-determined from the visual observations with the UVES spectrometer described in \citet{2007A&A...465..799L} and the oxygen abundances were re-determined from the H-band data described in \citep{2010A&A...509A..20R}. The UVES observations were carried out May-Aug 2003-2004 and the CRIRES observations were done May-Aug 2007-2008 and a summary of the observations and the S/N reached, is presented in Table \ref{tab:obslog}. The large optical extinction in the Bulge direction is the cause of the large differences in exposure times between the visual and the infrared observations. The extinction in the K band is a factor of 10 lower that in the V band \citep{1989ApJ...345..245C}.

In addition to these eight giants, K-band spectra from three K giants and two M giants in \citet{2008ApJ...679L..17C} (in turn from \citet{2006ApJ...651..491C}) have been re-analyzed. These stars are all in Baade's Window and were observed using the Phoenix spectrograph at Gemini-South \citep{1998SPIE.3354..810H}. For a complete description of these observations, see \citet{2008ApJ...679L..17C} and \citet{2006ApJ...651..491C}. 

\begin{table}[htp]
\caption{Summary of the observations with VLT/UVES and VLT/CRIRES.}
\begin{tabular}{l r r r c c c}
\hline
\hline
Star & \multicolumn{3}{c}{Total integration time} & \multicolumn{3}{c}{S/N\tablefootmark{a}}\\
     & \multicolumn{1}{c}{Visual} & \multicolumn{1}{c}{$H$} & \multicolumn{1}{c}{$K$} & \multicolumn{1}{c}{Visual} & \multicolumn{1}{c}{$H$} & \multicolumn{1}{c}{$K$}\\
\hline
\textrm{B3-b1} & 6h 10m  & 40m    & 52m    & 20 & 55 & 44\\
\textrm{B3-b7} & 6h 10m  & 1h 10m & 20m    & 38 & 31 & 37\\
\textrm{B3-b8} & 6h 10m  & 1h 04m & 1h 20m & 65 & 80 & 79\\
\textrm{B3-f3} & 11h 50m & ...    & 56m    & 31 & ... & 35\\
\textrm{BW-f6} & 6h 25m  & 1h 20m & 1h 20m & 34 & 46 & 38\\
\textrm{B6-b8} & 8h 30m  & 1h 04m & 1h 20m & 55 & 35 & 44\\
\textrm{B6-f1} & 5h 15m  & 32m    & 40m    & 75 & 33 & 28\\
\textrm{B6-f7} & 5h 15m  & 32m    & 1h 20m & 30 & 42 & 36\\
\hline
\end{tabular}
\label{tab:obslog}
\tablefoot{\\
\tablefootmark{a}{S/N per pixel as measured by the IDL-routine \texttt{der\textunderscore snr.pro}, see \href{http://www.stecf.org/software/ASTROsoft/DER\textunderscore SNR}{http://www.stecf.org/software/ASTROsoft/DER\textunderscore SNR}}
}
\end{table}

\section{Analysis}
The visual, as well as the infrared spectra, were analyzed using the software \texttt{Spectroscopy Made Easy}, SME \citep{1996A&AS..118..595V}. SME simultaneously fits a chosen number of parameters by fitting calculated synthetic spectra to parts of an observed spectrum using $\chi^2$-minimization. The parts, called line masks and continuum masks, mark regions with spectral lines of interest and points which SME should treat as continuum points. The latter are used if a linear rectification in predefined narrow windows of the already continuum-normalized observed spectrum is needed (see Section \ref{sec:cont}).

SME uses spherical symmetric, [$\alpha$/Fe]-enhanced, LTE MARCS-models. Within the Gaia-ESO collaboration \citep{2012Msngr.147...25G} it has also been developed to handle NLTE for many iron lines. We have no knowledge of estimated 3D-effects on the fluorine line used in the analysis for our stellar parameters, but \citet{2013ApJ...765...51L} have calculated 3D-corrections for more metal-poor stars showing that they are small.

\subsection{Stellar parameters} \label{sec:params}
In order to be consistent, we use SME in our analysis, both for our optical and infrared spectra. We have, thus, also redetermined the stellar parameters for our stars based on the method described in J\"onsson et al. (in prep.). In short, we determine all the stellar parameters ($T_\textrm{eff}$, $\log g$, [Fe/H], and  $\xi_\textrm{micro}$) simultaneously, with SME using a well-chosen line-list of weak, unblended $\ion{Fe}{i}$, $\ion{Fe}{ii}$, and $\ion{Ca}{i}$ lines and gravity-sensitive $\ion{Ca}{i}$-wings. All lines except some $\ion{Fe}{ii}$-lines have lab-measured oscillator strengths with excellent accuracy (according to the Gaia-ESO line-list categorization of Heiter et al. (in prep.)) and for all iron lines NLTE-corrections have been used. The resulting parameters are listed in Table \ref{tab:stellarparams} and are in agreement, within uncertainties, with the ones in \citet{2010A&A...509A..20R}.

In Table \ref{tab:stellarparams} we also list the stellar parameters used for the Bulge stars of \citet{2008ApJ...679L..17C} that we re-determine the fluorine abundance for \citep{2006ApJ...651..491C}. These stellar parameters are determined from a combination of photometry and IR spectroscopy which might lead to systematic differences to the stellar parameters of the B3-BW-B6 data set. Note also that the two M giants are cooler and have a lower surface gravity than the rest of the stars perhaps leading to systematic differences as well.

\begin{table}[htp]
\caption{Determined stellar parameters for the reference star Arcturus and our program stars. Also listed are the stellar parameters of the re-analyzed stars from \citet{2008ApJ...679L..17C}.}
\begin{tabular}{l c c r c c}
\hline
\hline
Star &  $T_\textrm{eff}$ & $\log g$ & [Fe/H]\tablefootmark{a} & [$\alpha$/Fe]\tablefootmark{b} & $\xi_\textrm{micro}$ \\
     & [K]               & (cgs)    &                   &      & [km\,s$^{-1}$]       \\
\hline
  Arcturus\tablefootmark{c} &  4262 &   1.62 &  $-0.63$ &   0.23 &   1.62 \\
\hline
     B3-b1 &  4372 &   1.11 &  $-1.03$ &   0.39 &   1.45 \\
     B3-b7 &  4261 &   1.86 &  $-0.09$ &   0.01 &   1.57 \\
     B3-b8 &  4282 &   1.67 &  $-0.75$ &   0.28 &   1.47 \\
     B3-f3 &  4573 &   2.55 &    0.19  &   0.00 &   1.76 \\
     BW-f6 &  4117 &   1.22 &  $-0.54$ &   0.20 &   1.70 \\
     B6-b8 &  3989 &   1.30 &  $-0.17$ &   0.05 &   1.46 \\
     B6-f1 &  4101 &   1.52 &  $-0.10$ &   0.02 &   1.65 \\
     B6-f7 &  4221 &   1.83 &  $-0.41$ &   0.14 &   1.63 \\
\hline
BMB 78\tablefootmark{d}     &  3600 &   0.8  & $-0.08$ & 0.01 & 2.5 \\
BMB 289\tablefootmark{d}    &  3375 &   0.4  & $-0.10$ & 0.02 & 3.0 \\
I-322\tablefootmark{d}      &  4250 &   1.5  & $-0.29$ & 0.10 & 2.0 \\
IV-072\tablefootmark{d}     &  4400 &   2.4  & $0.19$  & 0.00 & 2.2 \\
IV-329\tablefootmark{d}     &  4275 &   1.3  & $-0.57$ & 0.21 & 1.8 \\
\hline
\end{tabular}
\label{tab:stellarparams}
\tablefoot{\\
\tablefootmark{a}{ We use $\log \epsilon(Fe)_{\odot}=7.50$ \citep{2009ARA&A..47..481A}.\\}
\tablefootmark{b}{Following the SME MARCS model trends with [$\alpha$/Fe]=0.4 for [Fe/H]$<-1.0$, [$\alpha$/Fe]=0.0 for [Fe/H]$>0.0$, and linearly rising in-between.\\}
\tablefootmark{c}{Spectrum from the atlas by \citet{2000vnia.book.....H}.\\}
\tablefootmark{d}{Stellar parameters from \citet{2006ApJ...651..491C}.}
}
\end{table}

The uncertainties in our method of determining the stellar parameters from optical spectra and their dependence of S/N will be described in J\"onsson et al. (in prep.). In short we have degraded the Arcturus spectrum of \citet{2000vnia.book.....H} to different S/N and determined the stellar parameters for those spectra. The estimated uncertainties for the stars in this paper following this method are $\delta T_\textrm{eff} \la 70$ K, $\delta \log g \la 0.2$, $\delta$[Fe/H]$\la 0.1$, and  $\delta \xi_\textrm{micro} \la 0.1$.

\subsection{Line data}
All optical line data used in this paper has been collected and/or determined within the Gaia-ESO collaboration (Heiter et al., in prep). The infrared line data except for HF have been extracted from the VALD database \citep{1996A&AS..118..595V,1997BaltA...6..244R,1999A&AS..138..119K,2000BaltA...9..590K}. The line data of the [\ion{O}{i}]-line, the three \ion{Zr}{i}-lines, and the OH-lines used is listed in Table \ref{tab:linedata}. When it comes to the excitation energies and transition probabilities for HF we calculate them in Section \ref{sec:HF-Molecule}.

\begin{table}
\caption{Atomic and molecular data for the spectral lines used for O and Zr abundance determination.}
\begin{tabular}{c c c c c }
\hline
\hline
Element & Wavelength & $\chi_{\mathrm{exc}}$ & log(gf) & Refs.\\
\hline
$\ion{Zr}{i}$  & 6127.4400 & 0.154 & -1.060 & 1\\
$\ion{Zr}{i}$  & 6134.5500 & 0.000 & -1.280 & 1\\
$\ion{Zr}{i}$  & 6143.2000 & 0.071 & -1.100 & 1\\
$[\ion{O}{i}]$ & 6300.3038 & 0.000 & -9.715 & 2,3\\
OH             & 15558.021 & 0.304 & -5.309 & 4\\
OH             & 15560.241 & 0.304 & -5.309 & 4\\
OH             & 15565.838 & 3.663 & -4.830 & 4\\
OH             & 15565.961 & 2.783 & -4.700 & 4\\
OH             & 15568.780 & 0.299 & -5.270 & 4\\
OH             & 15572.083 & 0.300 & -5.270 & 4\\
\hline
\end{tabular}
\label{tab:linedata}
\tablebib{
(1) \citet{1981ApJ...248..867B};
(2) \citet{1966atp..book.....W};
(3) \citet{2000MNRAS.312..813S};
(4) \citet{1998JQSRT..59..453G}
}
\end{table}

\subsubsection{HF molecule} \label{sec:HF-Molecule}
The excitation energies and transition probabilities for HF have not been presented previously in a complete and comprehensive manner. The values of \citet{1992A&A...261..164J}, who cite private communications with Tipping, are often used. \citet{2011ApJ...729...40L}, \citet{2013ApJ...763...22D}, and \citet{2013AJ....146..153N}, however, use the excitation energy for the 23358.329 \AA-line from \citet{2000PhDT........16D}, in turn from private communications with Sauval, which differs from the Tipping value by 0.25 eV. As long as the excitation energies and partition functions are consistent they can both be used for abundance determinations if the corresponding partition function is used. Otherwise there will be an $\sim$0.3 dex difference in abundance just as \citet{2011ApJ...729...40L}, \citet{2013ApJ...763...22D}, and \citet{2013AJ....146..153N} show. Since it is unclear which partition function is used in most works it is difficult to compare the resulting abundance values. In this paper we intend to explicitly present which excitation energies, transition probabilities, and partition function we use so our data can be easily compared with coming studies.

The partition function is defined as: 
\begin{equation}
Q(T) =  \sum\limits_i g_i\cdot e^{-\chi_\mathrm{i}/kT}
\end{equation}
where $g_\mathrm{i}$ and $\chi_\mathrm{i}$ is the statistical weight and the excitation energy of level $i$. The consistent excitation energies have to be used when calculating the number density of a certain lower level for a transition: 
\begin{equation}
\label{eq:boltz}
\frac{n_\mathrm{lower}}{n_\mathrm{total}} = \frac{g_\mathrm{lower}}{Q(T)} \cdot e^{-\frac{\chi_\mathrm{lower}}{kT}} 
\end{equation}
The zero point energy of the levels used (which is an issue for molecules but not for atoms), must correspond to the one used to calculate the partition function. Thus, as long as there is not a mis-match, it does not matter which is used since the zero point energies can be factored out in Equation \ref{eq:boltz}.

\begin{figure}[htp]
\centering
\includegraphics[width=88mm]{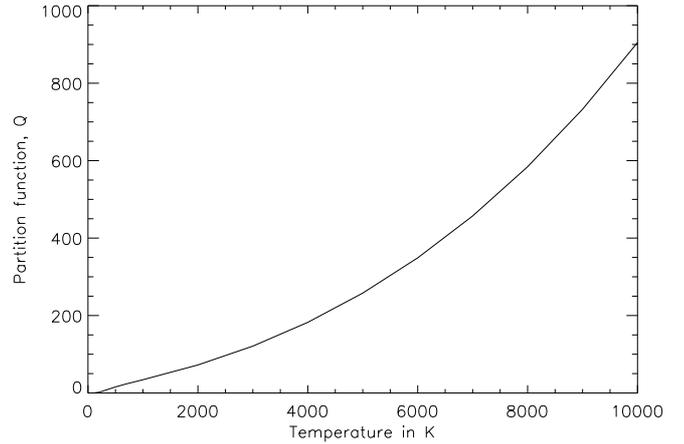}
\caption{Partition function of the HF molecule used in the MARCS code and SME for a relevant temperature range.}
\label{fig:part}
\end{figure}

We use the partition function from MARCS/BSYN and SME \citep[][and references therein]{2008A&A...486..951G}, which is an updated version of the one from \citet{1984ApJS...56..193S}. This partition function is shown in Equation \ref{eq:part} and in Figure \ref{fig:part}. 

\begin{multline}
\ln Q(T) = -360.5 + 222.4\cdot \ln T - 54.6 \cdot (\ln T)^2  + 6.69\cdot (\ln T)^3 \\ - 0.410 \cdot (\ln T)^4  + 0.00100\cdot (\ln T)^5
\label{eq:part}
\end{multline}

The dissociation energy used is the same as in \citet{1984ApJS...56..193S}: $D_0(HF)=5.869$ eV. We suspect that the 0.25 eV difference between different excitation energies used, comes from the Tipping-list using the dissociation energy of the energy potential, $D_e(HF)$, and not, like Sauval, the true energy required for dissociation, $D_0(HF)$. The former is larger than the latter due to the zero point of the energy of the lowest vibrational level. The difference is indeed 0.25 eV for HF  \citep{1991CPL...177..412Z}. \emph{We stress once again that it does not matter which energies are used as long as the consistent partition function is used.}

We have computed the HF line data from available molecular data consistent with the partition function and dissociation energy above. The excitation energies are calculated from the energy-level expression and coefficients of \citet{1994JMoSp.164..574L}. They fitted measured HF-line frequencies to the energy-level expression given by:
\begin{multline}
E(v,J)= T_v+B_vJ(J+1)-D_v[J(J+1)]^2+\\ H_v[J(J+1)]^3 +L_v[J(J+1)]^4,
\end{multline}
obtaining the rotational constants $T_v$, $B_v$, $D_v$, $H_v$, and $L_v$, especially for the vibrational states of interest for us, namely $v=0$ and $v=1$. These are provided in Table III of \citet{1994JMoSp.164..574L}. The calculated energy levels are good to $10^{-4}$\,cm$^{-1}$ or better. The excitation energies of the lower energy levels of the ro-vibrational lines of HF are presented in column 6 in Tables \ref{tab:gfr} and \ref{tab:gfp}.

\begin{table*}
\caption{HF line data$^{\mathrm{a}}$ for the {\it R branch} ($v'=1$ and $v''=0$).}
\begin{tabular}{l r r c c r c c c }
\hline
\hline
Line     & $J'$ & $J''$ & wavenumber  & wavelength             & $\chi_{exc}$ & $\chi_{exc}$ & $A_{v'J',v''J''}$ & $\log gf$  \\
         &      &       & $\sigma$    & $\lambda_\mathrm{air}$ &              &              &                   & \\              
R($J''$) &      &       & [cm$^{-1}$] & [\AA ]                 & [cm$^{-1}$]  & [eV]         & [s$^{-1}$]        &  \\
\hline
R(0)  &   1  &  0  &  4000.989  &    24987.001   &      0.00  & 0.000 & 63.42  &  -4.749 \\
R(1)  &   2  &  1  &  4038.962  &    24752.082   &     41.11  & 0.005 & 74.07  &  -4.468 \\
R(2)  &   3  &  2  &  4075.293  &    24531.418   &    123.28  & 0.015 & 77.02  &  -4.313 \\
R(3)  &   4  &  3  &  4109.936  &    24324.642   &    246.41  & 0.031 & 77.29  &  -4.209 \\
R(4)  &   5  &  4  &  4142.846  &    24131.413   &    410.35  & 0.051 & 76.26  &  -4.135 \\
R(5)  &   6  &  5  &  4173.979  &    23951.417   &    614.89  & 0.076 & 74.47  &  -4.079 \\
R(6)  &   7  &  6  &  4203.296  &    23784.365   &    859.78  & 0.107 & 72.19  &  -4.037 \\
R(7)  &   8  &  7  &  4230.756  &    23629.991   &   1144.73  & 0.142 & 69.54  &  -4.004 \\
R(8)  &   9  &  8  &  4256.322  &    23488.052   &   1469.37  & 0.182 & 66.64  &  -3.980 \\
R(9)  &  10  &  9  &  4279.960  &    23358.329   &   1833.32  & 0.227 & 63.53  &  -3.962 \\
R(10) &  11  & 10  &  4301.637  &    23240.623   &   2236.14  & 0.277 & 60.28  &  -3.950 \\
R(11) &  12  & 11  &  4321.321  &    23134.757   &   2677.32  & 0.332 & 56.91  &  -3.942 \\
R(12) &  13  & 12  &  4338.986  &    23040.574   &   3156.34  & 0.391 & 53.46  &  -3.940 \\
R(13) &  14  & 13  &  4354.604  &    22957.938   &   3672.62  & 0.455 & 49.97  &  -3.941 \\
R(14) &  15  & 14  &  4368.152  &    22886.733   &   4225.54  & 0.524 & 46.45  &  -3.947 \\
R(15) &  16  & 15  &  4379.608  &    22826.862   &   4814.44  & 0.597 & 42.94  &  -3.956 \\
R(16) &  17  & 16  &  4388.955  &    22778.249   &   5438.62  & 0.674 & 39.45  &  -3.969 \\
R(17) &  18  & 17  &  4396.176  &    22740.837   &   6097.33  & 0.756 & 36.01  &  -3.985 \\
R(18) &  19  & 18  &  4401.256  &    22714.589   &   6789.81  & 0.842 & 32.64  &  -4.007 \\
R(19) &  20  & 19  &  4404.184  &    22699.488   &   7515.25  & 0.932 & 29.36  &  -4.031 \\
R(20) &  21  & 20  &  4404.950  &    22695.539   &   8272.80  & 1.026 & 26.18  &  -4.061 \\
R(21) &  22  & 21  &  4403.548  &    22702.765   &   9061.58  & 1.123 & 23.13  &  -4.094 \\
R(22) &  23  & 22  &  4399.973  &    22721.213   &   9880.69  & 1.225 & 20.23  &  -4.133 \\
R(23) &  24  & 23  &  4394.221  &    22750.951   &  10729.19  & 1.330 & 17.47  &  -4.177 \\
R(24) &  25  & 24  &  4386.294  &    22792.072   &  11606.13  & 1.439 & 14.89  &  -4.228 \\
R(25) &  26  & 25  &  4376.191  &    22844.690   &  12510.51  & 1.551 & 12.49  &  -4.286 \\  
\hline
\end{tabular}
\label{tab:gfr}
\tablefoot{\\
\tablefootmark{a}{The consistent partition function is given in the text.}
}
\end{table*}

\begin{table*}
\caption{HF line data$^{\mathrm{a}}$ for the {\it P branch} ($v'=1$ and $v''=0$).}
\begin{tabular}{l r r c c r c c c }
\hline
\hline
Line     & $J'$ & $J''$ & Wavenumber  & Wavelength             & $\chi_{exc}$ & $\chi_{exc}$ & $A_{v'J',v''J''}$ & $\log gf$ \\
         &      &       & $\sigma$    & $\lambda_\mathrm{air}$ &              &              &                   & \\     
P($J''$) &      &       & [cm$^{-1}$] & [\AA ]                 & [cm$^{-1}$]  & [eV]         & [s$^{-1}$]        & \\
\hline
P(1)  &  0  &   1  &  3920.312   &  25501.219   &    41.11  & 0.005  &  199.3  &  -4.711 \\
P(2)  &  1  &   2  &  3877.707   &  25781.401   &   123.28  & 0.015  &  135.4  &  -4.393 \\
P(3)  &  2  &   3  &  3833.661   &  26077.610   &   246.41  & 0.031  &  123.9  &  -4.199 \\
P(4)  &  3  &   4  &  3788.227   &  26390.371   &   410.35  & 0.051  &  119.7  &  -4.058 \\
P(5)  &  4  &   5  &  3741.459   &  26720.249   &   614.89  & 0.076  &  117.8  &  -3.945 \\
P(6)  &  5  &   6  &  3693.412   &  27067.848   &   859.78  & 0.107  &  116.7  &  -3.850 \\
P(7)  &  6  &   7  &  3644.142   &  27433.815   &  1144.73  & 0.142  &  116.1  &  -3.769 \\
P(8)  &  7  &   8  &  3593.705   &  27818.843   &  1469.37  & 0.182  &  115.5  &  -3.696 \\
P(9)  &  8  &   9  &  3542.159   &  28223.674   &  1833.32  & 0.227  &  115.0  &  -3.632 \\
P(10) &  9  &  10  &  3489.559   &  28649.101   &  2236.14  & 0.277  &  114.4  &  -3.573 \\
P(11) & 10  &  11  &  3435.964   &  29095.974   &  2677.32  & 0.332  &  113.7  &  -3.518 \\
P(12) & 11  &  12  &  3381.432   &  29565.205   &  3156.34  & 0.391  &  112.8  &  -3.468 \\
P(13) & 12  &  13  &  3326.020   &  30057.770   &  3672.62  & 0.455  &  111.8  &  -3.422 \\
P(14) & 13  &  14  &  3269.785   &  30574.715   &  4225.54  & 0.524  &  110.6  &  -3.378 \\
P(15) & 14  &  15  &  3212.784   &  31117.163   &  4814.44  & 0.597  &  109.2  &  -3.337 \\
P(16) & 15  &  16  &  3155.075   &  31686.321   &  5438.62  & 0.674  &  107.7  &  -3.299 \\
P(17) & 16  &  17  &  3096.715   &  32283.483   &  6097.33  & 0.756  &  106.0  &  -3.262 \\
P(18) & 17  &  18  &  3037.758   &  32910.043   &  6789.81  & 0.842  &  104.1  &  -3.228 \\
P(19) & 18  &  19  &  2978.260   &  33567.501   &  7515.25  & 0.932  &  102.1  &  -3.195 \\
P(20) & 19  &  20  &  2918.275   &  34257.472   &  8272.80  & 1.026  &  99.94  &  -3.164 \\
P(21) & 20  &  21  &  2857.858   &  34981.701   &  9061.58  & 1.123  &  97.63  &  -3.134 \\
P(22) & 21  &  22  &  2797.061   &  35742.068   &  9880.69  & 1.225  &  95.18  &  -3.106 \\
P(23) & 22  &  23  &  2735.935   &  36540.611   & 10729.19  & 1.330  &  92.61  &  -3.078 \\
P(24) & 23  &  24  &  2674.531   &  37379.535   & 11606.13  & 1.439  &  89.94  &  -3.053 \\
P(25) & 24  &  25  &  2612.899   &  38261.231   & 12510.51  & 1.551  &  87.16  &  -3.028 \\
P(26) & 25  &  26  &  2551.086   &  39188.299   & 13441.33  & 1.667  &  84.30  &  -3.004 \\
\hline
\end{tabular}
\label{tab:gfp}
\tablefoot{\\
\tablefootmark{a}{The consistent partition function is given in the text.}
}
\end{table*}

From this we calculated the transition frequencies (and wavelengths) from the differences of the energy levels of the upper and lower level of the lines. The wavenumbers and wavelengths of the HF lines in the 22700-25000~\AA\, region (R branch, including the band head at 22700~\AA) and 25500-39200~\AA\,region (P branch) are given in columns 4 and 5, respectively, in the Tables \ref{tab:gfr} and \ref{tab:gfp}. The R-branch lines lie in the K band, whereas the P-branch lines originating from higher rotational levels lie in the L band.

We have also computed the HF ro-vibrational Einstein coefficients for spontaneous emission using the transition matix-element expansion coefficients given by \citet{1992JChPh..97.1734A}. They used accurate dipole-moment functions based on experimental data to find these coefficients:

\begin{equation}
A_{v'\rightarrow v''}(m)=\frac{64\pi^4}{3\,h}\nu^3\frac{\left|m\right|}{2J'+1}\left|R_{v'\rightarrow v''}(m)\right|^2,
\end{equation}

where $m = J''+1$ for the R branch, i.e. $J'\leftarrow (J''-1)$ and $m =-J''$ for the P branch, i.e. $J'\leftarrow (J''+1)$. The upper state is designed with a prime, $'$, and the lower state with a double prime, $''$. The transition matrix elements, $R_{v'\rightarrow v''}(m)$,  are given by  $R_{v'\rightarrow v''}(m) = a_0+a_1m+a_2m^2+a_3m^3$, where the expansion coefficients, $a_i$, are given in a Table VI in \citet{1992JChPh..97.1734A}. Finally, the $\log gf$ values are calculated from:  

\begin{equation}
\log (gf_{v'J',v''J''}) = \log \frac{(2J'+1) m_{e}c}{8\pi^2e^2}\nu^{-2}  \cdot A_{v'J',v''J''}
\end{equation}

see, for example, \citet{1983A&A...128..291L}. The calculated $\log (gf)$-values are given in column 9 in Tables \ref{tab:gfr} and \ref{tab:gfp}. \citet{1992JChPh..97.1734A} claim that these transition probabilities are reliable and well-established, and that they are in agreement with {\it ab initio} calculations of \citet{zemke1991}, providing confidence in the values.

\begin{figure}[htp]
\centering
\includegraphics[width=88mm]{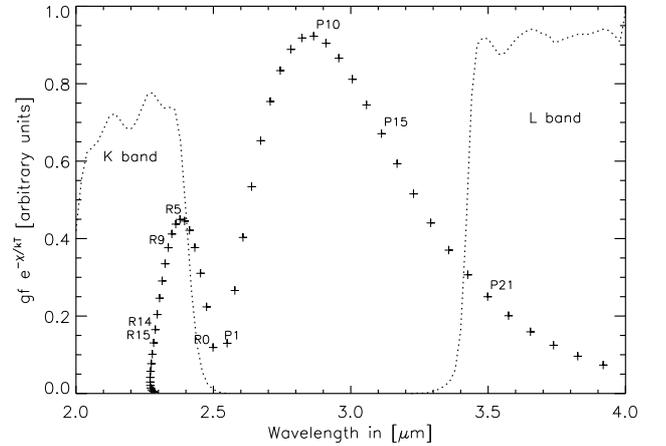}
\caption{Relative line strengths of the ro-vibrational HF lines (the R and P branches of the $v''=0$ to $v'=1$ band) given by $\mathrm{gf} \cdot e^{-\chi_\mathrm{exc}/kT}$ at $T=4000$~K. The K and L infrared transmission bands are indicated. }
\label{fig:linestrengths}
\end{figure}

To get an overview of which lines might be important for abundance determinations, we plot in Figure \ref{fig:linestrengths} the relative line strengths in the form of $\mathrm{gf} \cdot e^{-\chi_\mathrm{exc}/kT}$ at $T=4000$ K, a typical temperature of the line forming regions of a red giant. The R9-line used in this and many other works is marked together with some other lines. The equivalent widths of the lines for a typical model atmosphere, show in principle the same relative strengths.

\subsection{Stellar abundances}
All abundances for the B3-BW-B6 stars were determined using SME and the stellar parameters described in Table \ref{tab:stellarparams}. The abundances from the visual spectra were determined using the macro-turbulence determined simultaneously as the stellar parameters, but when determining the abundances from the IR-spectra the macro-turbulence was a global free parameter.

The uncertainties in the determined abundances from the uncertainties in the stellar parameters, see Section \ref{sec:params}, are given in Table \ref{tab:errors}.

\begin{table}[htp]
\caption{Uncertainties in the determined abundances due to uncertainties in the stellar parameters.}
\begin{tabular}{l c c c}
\hline
\hline
Uncertainty & $\Delta\log\epsilon(\mathrm{O})$ & $\Delta\log\epsilon(\mathrm{F})$ & $\Delta\log\epsilon(\mathrm{Zr})$\\
\hline
$\delta T_{\mathrm{eff}}=+70$ K      & $+0.12$  & $+0.15$  & $+0.14$\\ 
$\delta \log g= +0.2$                & $-0.02$  & $+0.01$  & $+0.02$\\
$\delta$[Fe/H]$=+0.1$                & $+0.06$  & $-0.03$  & $-0.01$\\ 
$\delta \xi_{\mathrm{micro}}= +0.1$  & $-0.01$  & $-0.01$  & $-0.01$\\ 
\hline
\end{tabular}
\label{tab:errors}
\end{table}

We note that all abundances are most sensitive to the temperature and that they all increase with higher temperature. This will mean that uncertainties, due to the uncertainties in the stellar parameters, in the ratios [F/O] and [Zr/F] used in Figures \ref{fig:koba_bulge} and \ref{fig:cunha_af} will be smaller than the quadratic addition of the two uncertainties. When it comes to the \emph{total} uncertainties in the abundances we also have to include the uncertainties in the continuum fitting around the O-, HF-, and Zr-lines used, but they are in most cases much smaller. Altogether we estimate the total uncertainties in the abundances to approximately 0.15 dex and in the abundance ratios to less than 0.1 dex.

The re-determination of the fluorine abundances for the sample of stars previously analyzed in \citet{2008ApJ...679L..17C} were done using the same LTE MARCS model atmospheres as the B3-BW-B6 stars, but using MOOG \citep{1973ApJ...184..839S} instead of SME. Our tests show that, using the same model atmosphere, SME and MOOG give the same result to a very good precision. For a discussion on the uncertainties of the stellar parameters and the abundances of the BMB-I-IV stars, see \citet{2006ApJ...651..491C} and \citet{2008ApJ...679L..17C}. In particular the most metal-poor star, IV-329, is challenging to analyze due to telluric lines.

\section{Results}
\begin{sidewaysfigure*}[htp]
\centering
\includegraphics[height=17cm]{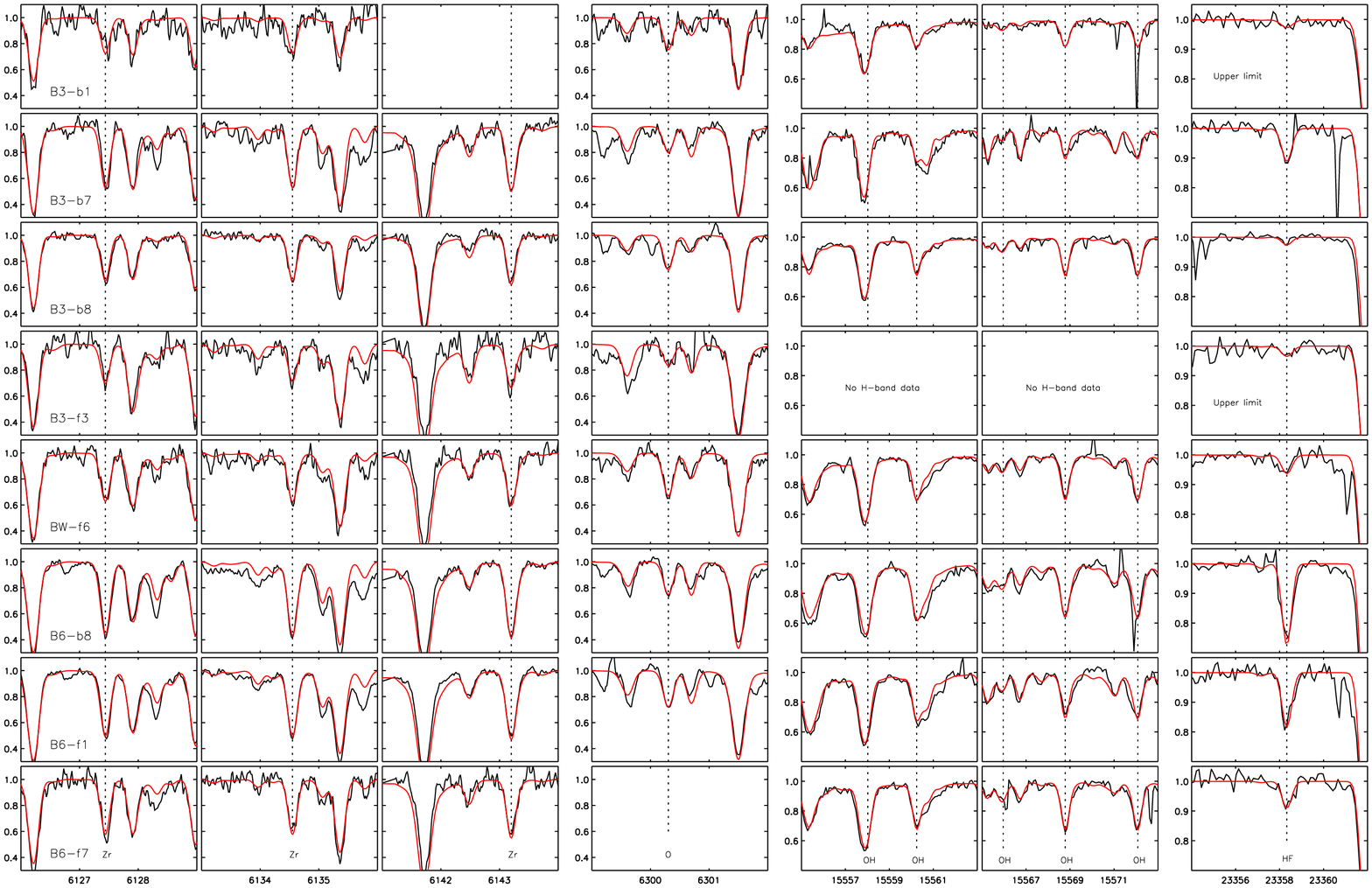}
\caption{Observed spectra in black and synthetic spectra in red for the B3-BW-B6 stars. The three Zr-lines, the [\ion{O}{i}]-line, the OH-lines, and the HF-line used are marked. The exact wavelengths are listed in Table \ref{tab:linedata}. The OH-lines with obvious cosmic hits have not been included in the fit. Note the different flux scales. An example of the HF-line in the re-analyzed BMB-I-IV can be found in \citet{2008ApJ...679L..17C}.}
\label{fig:spectra}
\end{sidewaysfigure*}

\begin{figure*}[htp]
\centering
\includegraphics[width=180mm]{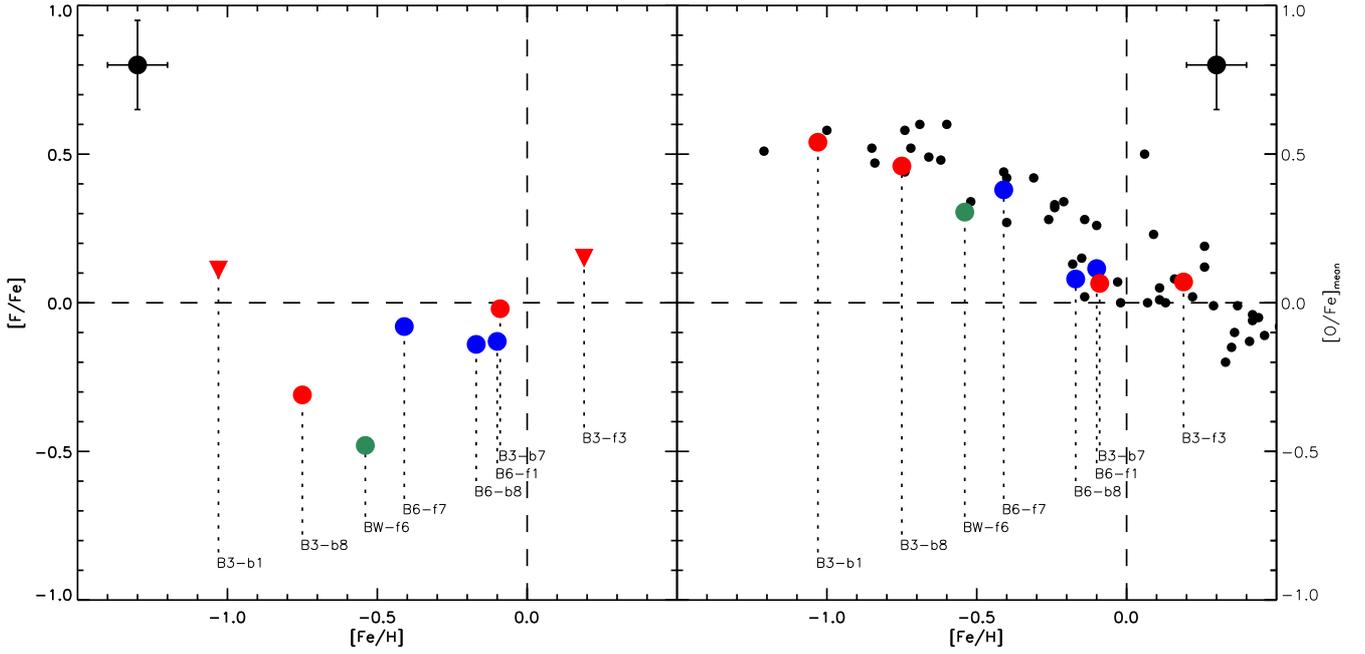}
\caption{[F/Fe] and [O/Fe] as functions of [Fe/H] for the B3-BW-B6 stars. The stars are color-coded as the corresponding fields in Figure~\ref{fig:bulge_fields}. The black dots are the micro lensed Bulge dwarfs of \citet{2013A&A...549A.147B} also marked in Figure~\ref{fig:bulge_fields}. A conservative estimation of the uncertainties are marked in the upper corners.}
\label{fig:feh}
\end{figure*}

\begin{figure*}[htp]
\centering
\includegraphics[width=180mm]{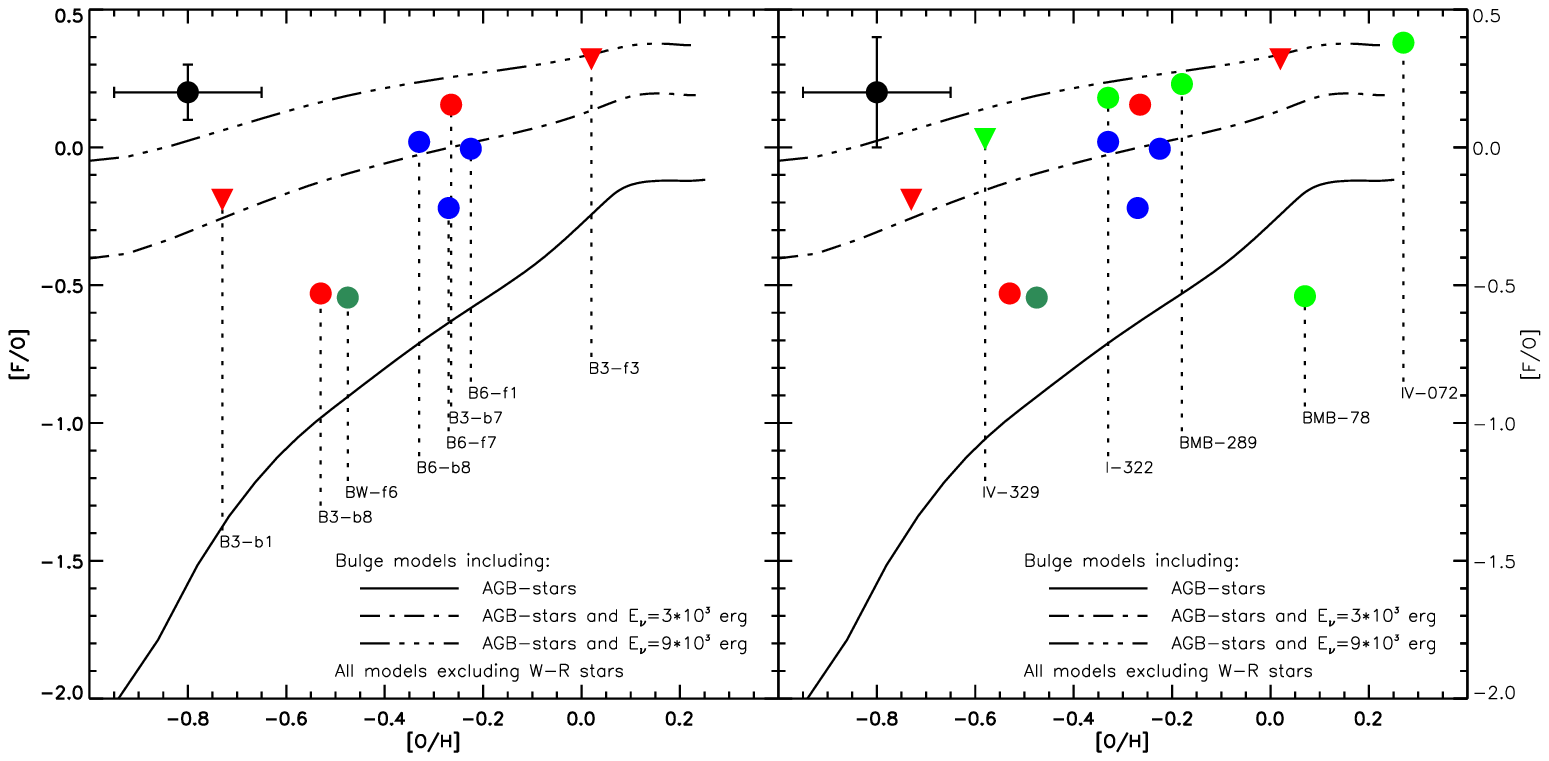}
\caption{Our fluorine abundances compared to the predictions of our Bulge models including AGB-stars, excluding \emph{and} including the $\nu$-process with two different energies, and excluding W-R stars. The abundances have been transformed to the scale of the models with $\log \epsilon(F)_{\odot}=4.56$ and $\log \epsilon(O)_{\odot}=8.93$ \citep{1989GeCoA..53..197A}. The stars are color-coded as the corresponding fields in Figure~\ref{fig:bulge_fields} with the BW-stars of \citet{2008ApJ...679L..17C} added in the right panel in a lighter green color. A conservative estimation of the uncertainties are marked in the upper left-hand corners. We note that the star BMB-78, still after the re-analysis, falls below the rest of the trend, see Section \ref{sec:discussion} for possible explanations.}
\label{fig:koba_bulge}
\end{figure*}

The part of the spectra containing the lines used in our investigation together with our best fitted synthetic spectra, are presented in Figure \ref{fig:spectra} and the resulting abundances are presented in Table~\ref{tab:results}.
In Figure \ref{fig:feh} we have plotted [F/Fe] and [O/Fe] as functions of [Fe/H] and in Figure \ref{fig:koba_bulge} we have plotted our abundances together with our chemical evolution models. 
The fluorine abundances derived here for the stars from \citet[][light-green circles]{2008ApJ...679L..17C} are systematically lower than those derived previously on account of the different excitation energies and partition functions used (as described in Section \ref{sec:HF-Molecule}), but also because in this study we use newer, alpha-enhanced stellar model atmospheres.

\begin{table*}[htp]
\caption{Determined abundances.}
\begin{tabular}{l c c c c c c c c}
\hline
\hline
Star & $\log \epsilon(\mathrm{O})_{[\ion{O}{i}]}$ & $\log \epsilon(\mathrm{O})_{\textrm{OH}}$ & $\log \epsilon(\mathrm{O})_{\textrm{mean}}$ & [O/Fe]$_{\textrm{mean}}$\tablefootmark{a} & $\log \epsilon(\mathrm{F})$ & [F/Fe]\tablefootmark{a} & $\log \epsilon(\mathrm{Zr})$ & [Zr/Fe]\tablefootmark{a}\\
\hline
  Arcturus\tablefootmark{b} &   8.58 &   8.47 &   8.52 &   0.47 &   3.75 &  -0.18 &  1.8 &  -0.11 \\
\hline
     B3-b1 &   8.11 &   8.29 &   8.20 &   0.54 &$\leq3.64$&$\leq0.11$&  2.0 &   0.53 \\
     B3-b7 &   8.68 &   8.65 &   8.66 &   0.07 &   4.45 &  -0.02 &  2.4 &  -0.02 \\
     B3-b8 &   8.41 &   8.39 &   8.40 &   0.46 &   3.50 &  -0.31 &  2.1 &   0.33 \\
     B3-f3 &   8.95 &    ... &   8.95 &   0.07 &$\leq4.90$&$\leq0.15$&  2.5 &  -0.24 \\
     BW-f6 &   8.51 &   8.40 &   8.45 &   0.31 &   3.54 &  -0.48 &  1.8 &  -0.22 \\
     B6-b8 &   8.54 &   8.66 &   8.60 &   0.08 &   4.25 &  -0.14 &  2.5 &   0.12 \\
     B6-f1 &   8.73 &   8.68 &   8.70 &   0.12 &   4.33 &  -0.13 &  2.3 &  -0.13 \\
     B6-f7 &    ... &   8.66 &   8.66 &   0.38 &   4.07 &  -0.08 &  2.2 &   0.05 \\
\hline
BMB-78 & ...  & 9.00\tablefootmark{c} & 9.00\tablefootmark{c} & 0.39\tablefootmark{c} &     4.09 &    -0.39 & ... & ... \\
BMB-289 & ... & 8.75\tablefootmark{c} & 8.75\tablefootmark{c} & 0.16\tablefootmark{c} &     4.61 &     0.15 & ... & ... \\
I-322 &  ...  & 8.60\tablefootmark{c} & 8.60\tablefootmark{c} & 0.20\tablefootmark{c} &     4.41 &     0.14 & ... & ... \\
IV-072 & ...  & 9.20\tablefootmark{c} & 9.20\tablefootmark{c} & 0.32\tablefootmark{c} &     5.21 &     0.46 & ... & ... \\
IV-329 & ...  & 8.35\tablefootmark{c} & 8.35\tablefootmark{c} & 0.23\tablefootmark{c} &$\leq4.01$&$\leq0.02$& ... & ... \\
\hline
\end{tabular}
\label{tab:results}
\tablefoot{\\
\tablefootmark{a}{Using solar abundances of $\log \epsilon(O)_{\odot}=8.69$, $\log \epsilon(F)_{\odot}=4.56$, $\log \epsilon(Fe)_{\odot}=7.50$, and $\log \epsilon(Zr)_{\odot}=2.58$ \citep{2009ARA&A..47..481A}.\\}
\tablefootmark{b}{Spectrum from the atlas by \citet{1995iaas.book.....H}.\\}
\tablefootmark{c}{From \citet{2006ApJ...651..491C}.}

}
\end{table*}

\section{Discussion}\label{sec:discussion}
\begin{figure}[htp]
\centering
\includegraphics[width=88mm]{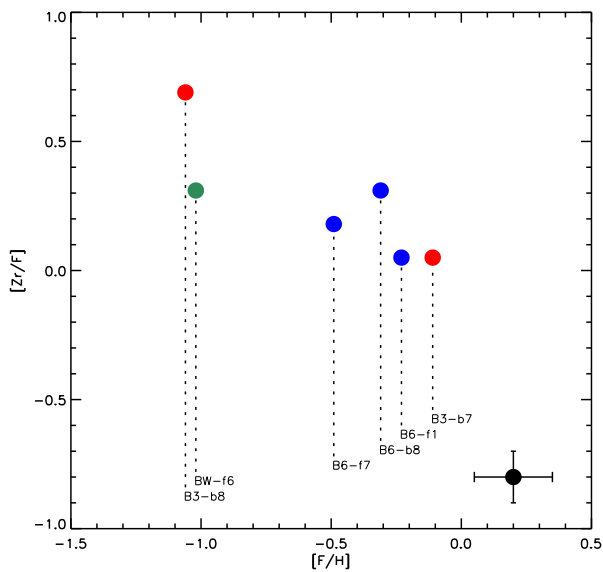}
\caption{Abundance ratios of fluorine and zirconium in our sample as a function of the solar normalized fluorine abundance. The negative slope alluded by this plot, suggests that W-R stars might be important for producing fluorine in the Bulge. The stars are color-coded as the corresponding fields in Figure~\ref{fig:bulge_fields} and a conservative estimation of the uncertainties are marked in the lower right-hand corner.}
\label{fig:cunha_af}
\end{figure}

From the color-coding in Figures \ref{fig:feh}-\ref{fig:cunha_af}, which designates the three different Bulge fields observed (see Figure \ref{fig:bulge_fields}), we are not able to trace any spatial variation of the fluorine abundance for the different fields. More stars in every field are needed in order to start discussing abundance trends. Therefore, in the following, we will discuss all our abundances as following a general Bulge-trend.

In the right panel of Figure \ref{fig:feh} we see the expected decline of oxygen with respect to [Fe/H] due to the large production of iron in SNe type Ia. Our trend follows closely the trend of the micro lensed Bulge stars of \citet{2013A&A...549A.147B}. In the left panel we do \emph{not} see a decline in [F/Fe] for the same range of [Fe/H], and if the upper limit of fluorine abundance in the star B3-b1 is ignored the trend in [F/Fe] increases for the same metallicity range. This would indicate that there must be a production site of fluorine (over-) compensating for the increase in iron around the same time scale as SN type I.

\citet{2011MNRAS.414.3231K} model the evolution of fluorine including `normal' nucleosynthesis in supernovae and synthesis in AGB-stars, excluding the $\nu$-process and yields from W-R-winds for different populations of the Galaxy, including the Bulge. In \citet{2011ApJ...739L..57K} they show models also including the $\nu$-process in SNe II, with two different neutrino energies, but only for the solar neighborhood. The two energies chosen are $E_\nu=3\times 10^{53}$\,erg (the energy estimated by \citet{1991NuPhA.527..663H} and best reproducing the values of \citet{2013ApJ...765...51L} in the solar neighborhood) and $E_\nu=9\times 10^{53}$\,erg (the theoretically largest possible value). In Figure \ref{fig:koba_bulge} we present for the first time the combination of the Bulge model as described in \citet{2011MNRAS.414.3231K} with the $\nu$-process as modeled in \citet{2011ApJ...739L..57K}. If we for the moment ignore the light-green values for the re-analyzed \citet{2008ApJ...679L..17C} stars, we find, just like \citet{2013ApJ...765...51L} do for the solar neighborhood, that the fluorine-oxygen abundance trend in the Bulge is \emph{best} described with the model including AGB-stars and $E_\nu=3\times 10^{53}$\,erg, but that the models do not to reproduce the trend of the lower-metallicity stars: there seem to be a larger slope in the observed data than in the models, see left panel of Figure \ref{fig:koba_bulge}. This might be due to that the $\nu$-process-contribution in the Bulge is more metallicity-dependent than in the models, or that its contribution in the Bulge is lower and another, more metallicity-dependent source (possibly W-R-stars) is needed.

When including the light-green stars from \citet{2008ApJ...679L..17C}, the trend becomes somewhat more scattered, and we treat their fluorine abundance for the star IV-329 as an upper limit.  At this star's temperature and metallicity, the HF line is only 4\% deep and may be affected significantly by imperfect telluric division. The recent discussion by \citet{2013A&A...560A..74D} notes that in low-metallicity stars where the HF R9 line becomes very weak, uncertain telluric-line removal at the 1-2\% level (which they maintain is typical) results in large fluorine abundance uncertainties. They suggest caution in interpreting fluorine abundances derived from such a weak HF R9 line. Further, since the figure shows [F/O] as a function of [O/H] the accuracy of the oxygen abundance plays a vital role in defining the trend. As an example of how sever the impact of the oxygen abundance is on the trend of Figure \ref{fig:koba_bulge} we note that decreasing the oxygen abundance of the peculiar star BMB-78 by 0.3 dex to better follow the B3-BW-B6-stars in Figure \ref{fig:feh} will shift it into the B3-BW-B6-trend in Figure \ref{fig:koba_bulge}. The oxygen-abundances of the \citet{2008ApJ...679L..17C} stars will be re-determined using newer, alpha-enhanced model atmospheres to see whether this will influence the [F/O]-trend amongst these stars in Figure \ref{fig:koba_bulge} (Cunha \& Smith, in prep). Presently we cannot rule out that the peculiar fluorine abundance of BMB-78 is a result of inhomogeneous chemical evolution in the Bulge (see \citet{2008ApJ...679L..17C} for further discussion on this). The other \citet{2008ApJ...679L..17C}-stars agree well with the B3-BW-B6 data set in spite of the systematic differences expected from different methods of determining the stellar parameters and the fact that BMB-289 is a M-giant, while the entire B3-BW-B6 data set is made up by K-giants.
 
To further investigate the possible need of W-R stars to explain the fluorine abundance in the Bulge, we have determined the abundance of the s-element zirconium mainly produced in low-mass AGB-stars \citep{2004ApJ...601..864T} and compared it to the abundance of fluorine, see Figure \ref{fig:cunha_af}. The negative slope in this plot, showing [Zr/F] as a function of [F/H], suggests that the most fluorine-rich stars have been enhanced in fluorine by a source \emph{not} producing zirconium, i.e. that the contribution from AGB-stars seems small. Thus, the additional sources needed might either be the $\nu$-process, W-R stars, or both. The $\nu$-process does not seem to be metal dependent in our [O/H]-range, while the W-R stars are, meaning that the slope is best explained with W-R stars. However, Zr is not exclusively produced in AGB-stars, but there is some minor r-process and weak s-process production in massive stars as well \citep{2004ApJ...601..864T,2011MNRAS.418..284B}. The contribution of these stars to the Bulge Zr-abundance is, as far as we know, not known. To evaluate this further modeling is needed. 

Concerning the role of W-R stars for the chemical evolution in the Bulge, it is of interest to compare to the discussion on the oxygen and magnesium trends in the Bulge \citep[for a review, see ][]{2013pss5.book..271R}. Since the metallicity-dependent, radiation-driven winds of W-R stars can be massive and cause the outer layers of the stars to be pealed off \citep{1992A&A...264..105M}, the contribution from these stars could explain the decline of oxygen in the Bulge \citep{2007ApJ...661.1152F,2008AJ....136..367M}.  At the same time these  massive stars could be an important formation site for carbon in the Galaxy \citep[see for example][]{1999A&A...342..426G,2010A&A...515A..68M}. \citet{2007ApJ...661.1152F} and \citet{2008AJ....136..367M} discuss the decline in [O/Mg] and [O/Fe] at solar metallicities and \citet{2009A&A...505..605C} the [C/O] versus [O/H] trends in the Bulge, and show that these are best fitted with models including massive star yields altered by metallicity-dependent winds, just as for W-R stars. Note, however, that \citet{2010A&A...513A..35A} and \citet{2010A&A...509A..20R} do not find the large increase in the  carbon abundance in the Bulge, which would have been expected if the W-R stars had played an important role. Thus, the question of the role of W-R stars in the Bulge is still open. It will, however, be able to be tested with more observations of the sort that already exists. Detailed modeling is needed and improved data may solve this issue \citep{2013pss5.book..271R}.

Since fluorine is produced from nitrogen in both AGB-stars and W-R stars, while it is produced from neon in the $\nu$-process it would be of interest to investigate the trend of F vs. N in the Bulge, but since our stellar sample is made up by giants it is hard to establish the `cosmic' nitrogen abundance to the needed accuracy due to newly produced nitrogen being dredged-up into the atmosphere of the star.

\section{Conclusion}
At low metallicity, our observed fluorine-oxygen abundance trend in the Bulge is lower than predicted in our Bulge model including the $\nu$-process, showing a steeper slope than the model. This might suggest a metal-dependent production source of fluorine. This source cannot be the $\nu$-process in SNe II because it is not metal-dependent over our metallicity range, and it cannot be AGB-stars because these produce s-elements at the same time as fluorine and would probably not give rise to the observed decline in [Zr/F] for increasing [F/H] (as shown in Figure \ref{fig:cunha_af}). Therefore our data corroborate the findings of \citet{2008ApJ...679L..17C} that W-R stars might be an important source of fluorine in the Bulge. To fully evaluate this we need galactic chemical evolution models that include full sets of yields of AGB stars, W-R stars, and supernova explosions.

We believe that some of the earlier reports of high fluorine abundances might be due to the use of mis-matching molecular data for the HF-molecule, but this has to be investigated. To help with this we have presented a HF line-list with a consistent partition function for lines in the K- and L-bands.

\begin{acknowledgements}
This research has been partly supported by the Royal Physiographic Society in Lund, Stiftelsen Walter Gyllenbergs fond and M\"arta och Erik Holmbergs donation. Also support from the Swedish Research Council, VR, project number 621-2008-4245, is acknowledged. N.R. is a Royal Swedish Academy of Sciences Research Fellow supported by a grant from the Knut and Alice Wallenberg Foundation. N.R. would like to thank the Aspen Center for Physics (and the NSF Grant \#1066293) for hospitality during the Bulge/Bar workshop in September 2011, at which part of this work was initiated. M.Z. acknowledges support  by Proyecto Fondecyt Regular 1110393, the BASAL  Center for  Astrophysics and  Associated Technologies PFB-06, and by the Chilean Ministry for the Economy, Development, and Tourism's Programa  Iniciativa Cient\'{i}fica Milenio  through grant P07-021-F, awarded to The Milky Way Millennium Nucleus. This publication made use of the SIMBAD database, operated at CDS, Strasbourg, France, NASA's Astrophysics Data System, and the VALD database, operated at Uppsala University, the Institute of Astronomy RAS in Moscow, and the University of Vienna.
\end{acknowledgements}

\bibliography{/Users/henrik/Documents/Bibliografi/papers.bib}
\bibliographystyle{aa}

\end{document}